\documentclass{aastex62}

\usepackage{amsmath}
\usepackage{float}

\graphicspath{{./}{figures/}}

\received{--}
\revised{--}
\accepted{--}
\submitjournal{ApJ}

\shorttitle{GALE}
\shortauthors{Stejko et al.}

\begin{document}

\title{Helioseismic Modeling of Background Flows}


\correspondingauthor{Andrey Stejko}
\email{ams226@njit.edu}

\author{Andrey M. Stejko}
\affiliation{New Jersey Institute of Technology\\
 323 Dr Martin Luther King Jr Blvd.\\
 Newark, NJ 07012, USA}

\author{Alexander G. Kosovichev}
\affiliation{New Jersey Institute of Technology\\
 323 Dr Martin Luther King Jr Blvd.\\
 Newark, NJ 07012, USA}

\author{Nagi N. Mansour}
\affiliation{University of Illinois at Urbana-Champaign\\
104 S Wright St.\\ 
Urbana, IL 61801, USA}

\begin{abstract}

\indent We present a 3-dimensional (3D) numerical solver of the linearized compressible Euler equations (GALE -- Global Acoustic Linearized Euler), used to model acoustic oscillations throughout the solar interior. The governing equations are solved in conservation form on a fully global spherical mesh ($0 \le \phi \le 2\pi$, $0 \le \theta \le \pi$, $0 \le r \le R_{\odot}$) over a background state generated by the standard Solar Model S. We implement an efficient pseudo-spectral computational method to calculate the contribution of the compressible material derivative dyad to internal velocity perturbations, computing oscillations over arbitrary 3D background velocity fields. This model offers a foundation for a ``forward-modeling'' approach, using helioseismology techniques to explore various regimes of internal mass flows. We demonstrate the efficacy of the numerical method presented in this paper by reproducing observed solar power spectra, showing rotational splitting due to differential rotation, and applying local helioseismology techniques to measure travel times created by a simple model of single-cell meridional circulation.

\end{abstract}

\keywords{Helioseismology, Computational Methods, Solar meridional circulation}


\section{Introduction}
\label{sec:intro}

\indent Since the first measurements of 5-minute solar oscillations nearly 60 years ago \citep{1962ApJ...135..474L, 1979Natur.282..591C} helioseismology has evolved into a rich and diverse field using surface observations to probe solar depths, inferring solar parameters and dynamics. Comprehensive overviews of helioseismology can be found in \citet{2002RvMP...74.1073C,2003LNP...599...31D,2005LRSP....2....6G}. The global nature of these oscillations has allowed for some of the most precise measurements of the solar interior that are currently available and have become an indispensable tool in measuring internal solar rotation \citep{1984Natur.310...22D,1998ApJ...505..390S} and meridional circulation \citep{1997Natur.390...52G,2004ApJ...603..776Z}.

\indent The nature of Solar rotation is a complex problem and carries with it important information on the dynamic structure of the Sun. Accurate modeling of internal rotation is vital to understanding effects that rapid rotation rates and angular momentum transport can play in internal mixing of elements and hence, the evolution of solar structure. Observations of the solar surface show a pattern of differential rotation \citep{1990ApJ...351..309S} which varies widely in its period at the equator ($\sim 24$ days) to near the poles ($\sim 30$ days). This pattern of differential rotation is mimicked throughout the convective interior (upper $30\%$ of the solar radius), with a slight maximum in the velocity of subsurface layers ($\sim 0.95R_{\odot}$). These rates begin to converge at the tachocline \citep{1996ApJ...469L..61K} where the differential rotation is coupled to a solid core rotating at rate of $\sim 430\text{ nHz}$. Even though helioseismology has made unprecedented strides in modeling interior solar dynamics, many unanswered questions remain. Inversion results near the poles show some discrepancies between the Global Oscillation Network Group (GONG) and the Michelson Doppler Imager (MDI) \citep{2002ApJ...567.1234S} data pipelines as well as newer HMI \citep{2011JPhCS.271a2061H} observations. There are also conflicting measurements of the solar core's rotation rate using low-degree modes ($l=1-4$) -- ranging from significantly lower rates (BiSON, \citet{1996MNRAS.283L..31C}) to much higher ones (IRIS, \citet{1996SoPh..166....1L}). Recent measurements of g-modes have also implied rotation rates more than twice as fast as previous estimates \citep{2017A&A...604A..40F}.
\\
\indent Perpendicular to global rotation, we see strong poleward flows ($20\text{ m s}^{-1}$) in each hemisphere \citep{1996ApJ...460.1027H}. These meridional mass flows operate as mechanisms that distribute angular momentum and magnetic flux throughout the convective interior \citep{2003ApJ...589..665H}. Techniques in local helioseismology use perturbations in acoustic travel times to observe these flows \citep{2005LRSP....2....6G}, however it becomes more difficult to clearly resolve structures at greater depths. There is still no consensus on the nature and location of the return flow of meridional circulation cells -- while commonly thought to sit in the relatively deep region at the base of the tachocline \citep{1999_giles}, new techniques and measurements have begun to question this assumption \citep{2007AN....328.1009M, 2012ApJ...760...84H}. The structure of these circulation cells has also been put to doubt, with recent measurements inferring more than just a single cell per hemisphere \citep{2013ApJ...774L..29Z}, cf. \citet{2020Sci...368.1469G}.
\\
\indent Through persistent uncertainties on the nature of solar structure, forward-modeling offers an important avenue for the validation and systematic analysis of the helioseismology techniques used to infer internal solar parameters. Numerical acoustic solar models have been instrumental in testing these techniques under approximate solar conditions; since the seminal work of \citet{2003ESASP.517..319J}, simulating subsurface sound speed perturbations \citep{2007ApJ...666..547P,2014ApJ...785...40P}, 3D cartesian models of the solar atmosphere have been employed to validate local time-distance measurements in regimes of convection \citep{2007ApJ...669.1395B} and magnetic fields \citep{2008SoPh..251..291C,2009ApJ...694..411K,2009ApJ...694..573P,2009ASPC..416...61P,2016ApJ...829...67F}. The limitations of the computational domains, needed to probe deeper structures, led to the development of global spherical models \citep{2006ApJ...648.1268H,2007ApJ...664.1234H,2008ApJ...689.1373H,2013ApJ...762..132H,2015A&A...577A.145P,2017A&A...600A..35G} -- providing key validations of inversion techniques and inferred observations of sound speed perturbations and background flow structures in the hydrodynamic regime. These models have provided a strong basis for evaluating mean flow structures on the Sun - providing support \citep{2013ApJ...762..132H} for the inference of a double-cell \citep{2013ApJ...774L..29Z,2019_chen} meridional velocity profiles in HMI observations; however, with recent inferences that reaffirm the single-cell structure \citep{2020Sci...368.1469G} on MDI and GONG data, it becomes clear that a detailed systematic investigation of these techniques and the limits of resolving internal structures is required. In order to address some of these issues we demonstrate the efficacy of a new pseudo-spectral global acoustic algorithm -- developed to quickly and flexibly compute stochastically excited oscillations over wide varieties of static or dynamic 3-dimensional background velocity fields. Such fully global solar simulations have the potential to offer insights on global effects of complex hydrodynamic structures in the solar interior -- from modeling meridional circulation and center-to-limb effects \citep{2019_chen} to exploring the information that g-modes carry about the rotating solar core. The computational techniques presented in this paper will also become the basis for adding linear contributions of global magnetic field terms in future work.\\
\indent This paper is organized as follows: in \S\ref{sec:model} we describe the mathematical background and computational set-up of the model. In \S\ref{sec:nume} we detail the specific numerical methods we used to compute our governing equations. In \S\ref{sec:valid} we reproduce the solar p-mode spectrum and frequency splitting due to differential rotation. We continue our validation by reproducing measurements of a simple single-cell model of meridional circulation using local helioseismology techniques in \S\ref{sec:mer}. Finally, we discuss future plans for the model and state our concluding remarks in \S\ref{sec:conclusions}. 

\section{Model Description} \label{sec:model}
\subsection{Mathematical Background}
\label{sec:mathbg}

In this section we present the mathematical formulation and computational set-up for the GALE (Global Acoustic Linearized Euler) code -- a 3-dimensional numerical solver of the linearized compressible Euler equations, used to model acoustic oscillations throughout the solar interior. For the sake of simplicity we assume the adiabatic approximation, where the time-scale of heat transfer is much smaller than the period of oscillations and is therefore neglected in the conservation of energy \citep{christensen14}. The governing equations are solved on a fully global spherical mesh, $0 \le \phi \le 2\pi$, $0 \le \theta \le \pi$, $0 \le r \le R_{\odot}$, in the Cauchy conservation form; they are enumerated as follows:

\begin{equation}\label{eq:gov1}
  \dfrac{\partial \rho'}{\partial t} + \Upsilon' = 0 \ ,
\end{equation}
\begin{equation}\label{eq:gov2}
  \dfrac{\partial\Upsilon'}{\partial t} + \boldsymbol{\nabla}:\left(\mathbf{m}'\tilde{\mathbf{u}} + \tilde{\rho}\tilde{\mathbf{u}}\mathbf{u}'\right) = -\nabla^{2}\left(p'\right) - \nabla\cdot\left(\rho'\tilde{g}_{r}\mathbf{\hat{r}}\right) + \nabla\cdot S\mathbf{\hat{r}} \ ,
\end{equation}
\begin{equation}\label{eq:gov3}
  \dfrac{\partial p'}{\partial t} = - \dfrac{\Gamma_{1}\tilde{p}}{\tilde{\rho}}\left(\nabla\cdot\tilde{\rho}\mathbf{u}'+ \rho'\nabla\cdot\tilde{\mathbf{u}} - 
  \dfrac{p'}{\tilde{p}}\tilde{\mathbf{u}}\cdot\nabla\tilde{\rho} + \tilde{\rho} \mathbf{u}'\cdot\dfrac{N^{2}}{g}\mathbf{\hat{r}}\right) \ .
\end{equation}

In Eqns. (\ref{eq:gov1})-(\ref{eq:gov3}), we solve for oscillations in the potential field ($\phi$) independently, and solenoidal contributions are discarded, here $\Upsilon$ is defined as the divergence of the momentum field $\mathbf{m}$ ($\Upsilon = \nabla\cdot\mathbf{m} = \nabla\cdot\rho\mathbf{u} = \nabla^{2}\phi$). $\rho$, $u$, $p$, $g$ are the density, velocity, pressure and gravity terms respectively. $\Gamma_1$ is the adiabatic ratio, constant in time in the adiabatic approximation. The full derivation for current form of the energy equation (Eq. \ref{eq:gov3}) can be found in Appendix \ref{sec:eeq}. The governing equations are linearized - split into a base flow (tilde) and a perturbation from that base flow (prime), and only the $1^{\text{st}}$ order correlation terms are considered; the base values are derived from a theoretical background state – the standard solar model S \citep{1996Sci...272.1286C}. $N^{2}$ is the Brunt-V\"ais\"al\"a frequency where 
\begin{equation}
\label{eq:BVf}
    N^{2} = g\left(\dfrac{1}{\Gamma_1}\dfrac{\partial\ln p}{\partial r} - \dfrac{\partial\ln\rho}{\partial r}\right) \ .
\end{equation}

\noindent Our set of linear equations become unconditionally unstable in regions where this term is negative ($N^{2} < 0$). These instabilities can be avoided by deriving the conservation of energy (Eq. \ref{eq:gov3}) as a function of the Brunt-V\"ais\"al\"a frequency, allowing us to set slightly negative values in the convection zone to zero directly, without the need to alter background profiles of pressure, density or the adiabatic ratio ($\Gamma_{1}$) as in similar convectively stable models of \citet{2006ApJ...648.1268H,2007ApJ...666..547P,2008ApJ...689.1373H}. Altering the algorithm to maintain stability may introduce small deviations from the original model \citep{2014SoPh..289.1919P}, but is accurate enough to be used in testing helioseismology techniques throughout the convection zone.  \\
\indent The formulation of the source function ($S\mathbf{\hat{r}}$) is similar to that of \citet{2006ApJ...648.1268H}, where a thin shallow layer bounded by the model surface ($R_{\odot}$) simulates the source of solar oscillations, see \citet{1994ApJ...424..466G}, \citet{2001ApJ...546..585S}. This function is modeled as a radial Gaussian with a standard deviation of $\sigma = 0.0001 R_{\odot}\sim 69.6 \text{ km}$ and centered at $\mu = 0.9995 R_{\odot}$ -- simulating radial force perturbations from the solar surface. The full time-dependent spectrum of the source is generated in frequency space with a Gaussian function centered at $\mu = 3.2\text{ mHz}$ and with a standard deviation of $\sigma = 1\text{ mHz}$ -- simulating the power peak of observed solar oscillations. To mimic a stochastic excitation of our modes \citep{1984PhDT........34W}, we multiply our temporal Gaussian function by a set of random numbers at each frequency interval ($f_{s} = 1/\Delta t$); subsequently applying a Fourier transform to produce a randomized oscillating function for our source, of which a unique one is created for every harmonic degree and azimuthal order ($l,m$) in our spherical harmonic decomposition. 

\indent The material derivative ($\boldsymbol{\nabla}:\left(\mathbf{m}'\tilde{\mathbf{u}} + \tilde{\rho}\tilde{\mathbf{u}}\mathbf{u}'\right)$) is solved in its conservation form in order to fully account for effects that background velocities exert on the conservation of momentum (Eq. \ref{eq:gov2}) in the acoustic regime. This is an important requirement for properly simulating effects of differential rotation and meridional circulation on the frequency of solar oscillations. 

\subsection{Radial Grid}\label{sec:radgrid}

\indent The radial grid is spaced evenly with respect to acoustic travel time ($\int 1/c_{s} dr$) in order to compute acoustic oscillations across the large variations in sound speed ($c_{s}$) throughout the model. While this grid is effective at capturing effects throughout most of the model interior, as we move towards the surface ($r > 0.99 R_{\odot}$) pressure and density scale heights begin to drop off faster than sound speed. In order to resolve the effect of the Brunt-V\"ais\"al\"a frequency ($N^{2}$), we switch to a logarithmic pressure grid spaced evenly in $ln(p)$ \citep{2006ApJ...648.1268H}. Above the model surface ($r > R_{\odot}$), we implement a thin isothermal buffer layer with constant grid spacing up to $r = 1.001 R_{\odot}$.

\subsection{Boundary Conditions}

\indent The boundary layers are solved as simple reflective walls with a zero velocity perturbation condition ($\mathbf{u}'=0$). To avoid non-physical surface reflections from affecting our solution, we implement a buffer layer by placing a damping factor ($\sigma$) into our governing equations:

\begin{align}\label{eq:govdamp}
  \dfrac{\partial \rho'}{\partial t} = -\Upsilon' - \sigma\rho' \ , &&
  \dfrac{\partial\Upsilon'}{\partial t} = -\nabla^{2}p' + \mathcal{Y} - \sigma\Upsilon' \ , &&
  \dfrac{\partial p'}{\partial t} = \mathcal{P}- \sigma p' \ .
\end{align}

To avoid precision errors from our damping term, it is computed implicitly in our time-discretization scheme using the integrating factor method. Damping is initiated at the model surface ($R_{\odot}$) and steadily increased into the atmospheric layers, mimicking the escape of acoustic oscillations above the cutoff frequency ($>5$ mHz).

\section{Numerical Method}\label{sec:nume}

\indent The time-discretization scheme used in governing Eqns. (\ref{eq:gov1})-(\ref{eq:gov3}) is the $2^{\text{nd}}$ order accurate ($\mathcal{O}(h^{2})$) Adams-Bashforth method -- used to compute the effect of external contributions of background fields to our conservation of momentum and conservation of energy (Eqns. \ref{eq:gov2} - \ref{eq:gov3}) considerations, denoted by $\mathcal{Y}$ and $\mathcal{P}$ in Eq. (\ref{eq:govdamp}) respectively. To advance our continuity and conservation of momentum forward in time ($\rho'(\Upsilon'), \Upsilon'(p')$), (Eqns. \ref{eq:gov1} - \ref{eq:gov2}), we use the implicit $1^{\text{st}}$ order accurate ($\mathcal{O}(h)$) backward Euler method -- helping to maintain the stability of our solution. \\
\indent Spatial differentiation in the radial direction employs the $2^{\text{nd}}$ order accurate central finite-difference scheme (Eq. \ref{eq:poissonRD}) when computing the radial component of the Laplacian ($\nabla^{2}$) in Eq. \ref{eq:gov2}:\\

\begin{equation}
\label{eq:poissonRD}
  \frac{\partial}{\partial r}\left( r^{2}\frac{\partial f}{\partial r}\right) = \dfrac{2}{\Delta r_{k}+\Delta r_{k+1}} \left( r^{2}_{k+0.5}\dfrac{f_{k-1} - f_{k}}{\Delta r_{k+1}} - r^{2}_{k-0.5}\dfrac{f_{k} - f_{k-1}}{\Delta r_{k}} \right) \ ,
\end{equation}
\noindent where subscript $k$ denotes separate radial mesh points and $\Delta r_{k}$ represents the radial distance between points $k-1$ and $k$. The computation of the $1^{\text{st}}$ radial derivative also employs a $2^\text{nd}$ order accurate polynomial central finite-difference scheme (Fig. \ref{eq:polyRD}).
\begin{equation}
\label{eq:polyRD}
  \dfrac{\partial f}{\partial r} =  -\dfrac{\Delta r_{k+1}}{\Delta r_{k}(\Delta r_{k+1} + \Delta r_{k})}f_{k-1} + \dfrac{\Delta r_{k+1} - \Delta r_{k}}{\Delta r_{k+1}\Delta r_{k}}f_{k} + \dfrac{\Delta r_{k}}{\Delta r_{k+1}(\Delta r_{k+1} + \Delta r_{k})}f_{k+1}  \ .
\end{equation}
\noindent The precision of these techniques and their agreement with theoretical calculations of eigenmodes is quite high (Fig. \ref{fig:l-nu}) -- serving well to validate the computational techniques used in this algorithm; higher order accurate schemes may improve our results, but in the large variations in our non-uniform radial regime, many of the improvements that we expect may be greatly diminished.\\
\indent Differentiation tangent to the sphere is done in the pseudo-spectral regime through spherical harmonic decomposition using the Libsharp spherical harmonic library \citep{2013A&A...554A.112R}. Tangential resolution is defined by the maximum allowable quantum number ($l_{max}(r)$) at each radial point. This value controls the mesh size of our Gauss-Legendre grid, containing $N_{\phi} = 3l_{max}$ azimuthal grid points and $N_{\theta} = 3l_{max}/2$ latitudinal grid points. These values are chosen to avoid aliasing during spherical harmonic decomposition. The azimuthal mesh points ($N_{\phi}$) are spaced at even intervals between $0 < \phi < 2\pi$, while the latitudinal points ($N_{\theta}$) are placed at the roots of the corresponding Legendre Polynomial between $0 < \theta < \pi$. To avoid oversampling at high latitudes the Libsharp library natively implements polar optimisation, or the ``reduced Gauss-Legendre grid". \\
\indent To compute the tangential components of the divergence we use vector and tensor spherical harmonic bases (VSH and TSH) to calculate our terms spectrally, as defined in \S\ref{sec:vsh} and \S\ref{sec:tsh} respectively.

\subsection{Vector Spherical Harmonics}\label{sec:vsh}

\indent Using the ``pure-spin" vector spherical harmonic (VSH) components defined by \citet{1980RvMP...52..299T}, we can expand an arbitrary vector field (E) into the following linearly independent basis.

\begin{equation}\label{eq:vshe}
  \mathbf{E} = \sum\limits_{m=-lmax}^{lmax}\sum\limits_{l=|m|}^{lmax} \left( {E}^{r}_{lm}(r)\mathbf{Y_{lm}} + {E}^{(1)}_{lm}(r) \mathbf{\Psi_{lm}} + {E}^{(2)}_{lm}(r) \mathbf{\Phi_{lm}} \right) \ ,
\end{equation}

\noindent where $E^{r}_{lm}$ is the radial vector component, and $E^{(1)}_{lm}$, $E^{(2)}_{lm}$ are components tangential to the surface of the 2-sphere. Transformations between the spherical coordinate basis and our VSH basis is achieved by a set of functions defined in \citet{2010JCoPh.229..399N}, and computed using recurrence relations of the spherical harmonic ($Y_{lm}$). We define the divergence of our vector field ($\mathbf{E}$ in Eq. \ref{eq:vshe}) in our new basis with Eq. (\ref{eq:vshd}).

\begin{equation}\label{eq:vshd}
\nabla \cdot \mathbf{E} = \sum\limits_{m=-lmax}^{lmax}\sum\limits_{l=|m|}^{lmax} \left( \dfrac{d E^{r}_{lm}}{dr} + \dfrac{2}{r} E^{r}_{lm}-\dfrac{l(l+1)}{r} E^{(1)}_{lm} \right)Y_{lm}
\end{equation}

\indent The radial derivative is computed using the finite-difference method described in \S\ref{sec:nume}.

\subsection{Tensor Spherical Harmonics}\label{sec:tsh}

\indent The tensor spherical harmonic (TSH) basis is built on groups 0 and 2 of the irreducible representation of rotation in SO(3) ($\mathcal{D}^{0}$, $\mathcal{D}^{2}$ where $t = \mathcal{D}^{0} + \mathcal{D}^{1} + \mathcal{D}^{2}$), which correspond to the trace and the symmetric traceless tensor respectively \citep{mathews62}. This basis is coupled with spin-0 and spin-2 spherical harmonics to form 6 irreducible representations of a symmetric tensor, which can be rearranged into the ``pure-spin tensor harmonics" defined by \citet{1970JMP....11.2203Z} and \citet{1980RvMP...52..299T}. We can use these components to expand an arbitrary symmetric tensor ($\mathbf{t}$) into the following orthogonal basis \citep{2010JCoPh.229..399N}.

\begin{equation}\label{eq:tshe}
  \mathbf{t} = \sum\limits_{m=-lmax}^{lmax}\sum\limits_{l=|m|}^{lmax} \left( L^{0}_{lm}\mathbf{T^{L_{0}}_{lm}} + E^{1}_{lm}\mathbf{T^{E_{1}}_{lm}} + B^{1}_{lm}\mathbf{T^{B_{1}}_{lm}} + T^{0}_{lm}\mathbf{T^{T_{0}}_{lm}} + E^{2}_{lm}\mathbf{T^{E_{2}}_{lm}} + B^{2}_{lm}\mathbf{T^{B_{2}}_{lm}} \right) \ ,
\end{equation}

\noindent where $L^{0}_{lm}$ is the fully radial component ($t^{rr}$) and $T^{0}_{lm}$ is the transverse ($t^{\theta\theta}$, $t^{\phi\phi}$) portion of the trace. $E^{1}_{lm}$ and $B^{1}_{lm}$ represent the mixed radial/transverse components ($t^{r\theta}$, $t^{r\theta}$), and $E^{2}_{lm}$ and $B^{2}_{lm}$ are symmetric transverse traceless components. We can use this basis to solve for the divergence of a tensor in spherical coordinates, where

\begin{align}\label{eq:tshd}
    \nabla\cdot\mathbf{t}= &
  \begin{cases}
    \left[\nabla\cdot\mathbf{t}\right]^{r}_{lm}Y_{lm} =& \sum\limits_{m=-lmax}^{lmax}\sum\limits_{l=|m|}^{lmax} \left[\dfrac{1}{r^{2}}\dfrac{\partial}{\partial r}\left(r^{2}L^{0}_{lm}\right) - \dfrac{1}{r}\left(l(l+1)E^{1}_{lm} + T^{0}_{lm}\right)\right]Y_{lm} \ ,\\
    \left[\nabla\cdot\mathbf{t}\right]^{(1)}_{lm}Y_{lm} =& \sum\limits_{m=-lmax}^{lmax}\sum\limits_{l=|m|}^{lmax} -l(l+1)\left[\dfrac{1}{r^{3}}\dfrac{\partial}{\partial r}\left(r^{3}E^{1}_{lm}\right) + \dfrac{1}{r}\left(\dfrac{T^{0}_{lm}}{2}-(l-1)(l+2)E^{2}_{lm}\right)\right]Y_{lm} \ ,\\
    \left[\nabla\cdot\mathbf{t}\right]^{(2)}_{lm}Y_{lm} =& \sum\limits_{m=-lmax}^{lmax}\sum\limits_{l=|m|}^{lmax} -l(l+1)\left[\dfrac{1}{r^{3}}\dfrac{\partial}{\partial r}\left(r^{3}B^{1}_{lm}\right) - \dfrac{(l-1)(l+2)}{r}B^{2}_{lm}\right]Y_{lm} \ .
  \end{cases}
\end{align}

\noindent Coordinate transformations between the spherical, VSH and TSH bases is done using relations defined in \citet{2010JCoPh.229..399N}. We can plug the results of this computation into Eq. (\ref{eq:vshd}) to compute the divergence dyad ($\boldsymbol{\nabla:}$) in our material derivative (Eq. \ref{eq:gov2}).

\section{Model Validation}\label{sec:valid}

\indent In order to validate the mathematical set-up (\S\ref{sec:model}) and computational techniques (\S\ref{sec:nume}) described in this paper, we start by matching the generated power spectrum from our model with a theoretical prediction computed using the standard solar model S (\S\ref{sec:val_pow}). We test the computation of our material derivative (Eq. \ref{eq:gov2}) in two simple regimes of background velocity flows: differential rotation (\S\ref{sec:rot}) and a single-cell model of meridional circulation (\S\ref{sec:mer}). \\
\indent The parameters used in this validation are as follows: the surface resolution of the model is set by a spherical harmonic degree of $l_{max} = 200$, corresponding to $N_{\phi} = 600$ longitudinal and $N_{\theta} = 450$ latitudinal mesh points. These values are chosen in order to fully resolve the acoustic modes that intersect the base of the tachocline (a depth of $\sim 0.70 R_{\odot}$) -- letting us infer flows from any point in the convective interior using local helioseismology techniques \citep{2005LRSP....2....6G}. Our simulation is run for a period of 65 hours model time. This is too short to properly resolve flow velocities on the order of those seen in meridional circulation ($\sim 14\text{ m s}^{-1}$). In order to achieve a desirable signal-to-noise ratio in our validation of meridional circulation measurements, we increase the magnitude of our meridional velocity by a factor of $36$, to a maximum of $500\text{ m s}^{-1}$ -- simulating an SNR that would be observed over $\sim 9.5\text{ yrs}$, approaching the decade minimum that is estimated to be needed in order to resolve deep meridional flows \citep{2008ApJ...689L.161B}.

\subsection{Power Spectrum}\label{sec:val_pow}

\indent Acoustic oscillations on the solar surface, can be decomposed into eigenmodes of radial velocity ($u'_{r}\mathbf{\hat{r}}$), representing standing waves throughout the solar interior. These modes can be conceptualized as a combination of the scalar spherical harmonic ($Y_{lm}$) with a frequency dependent radial order ($\xi_n$),

\begin{equation}
u'_{r}(r,\theta,\phi,t) = \sum\limits_{n,l}\sum\limits_{m=-l}^{l} \xi_{n}(r) Y_{lm}(\theta,\phi)e^{i\omega t} \ .
\end{equation}

\begin{figure}[H]
\center
\includegraphics[width=6.2cm]{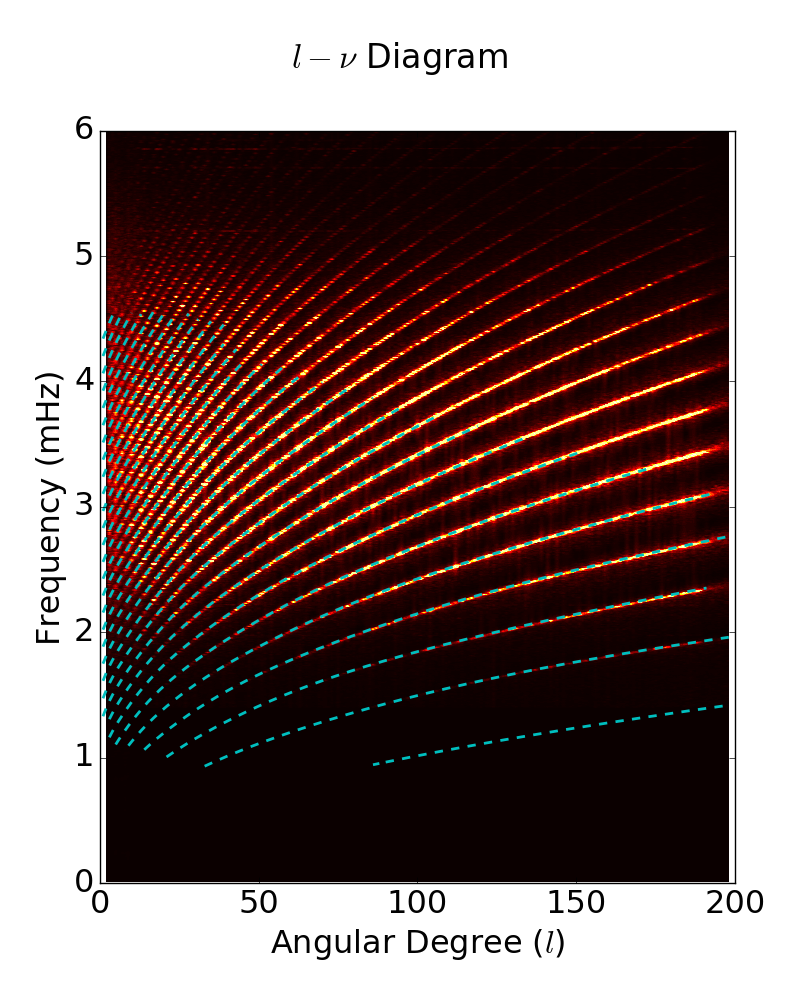}
\caption{An $l-\nu$ diagram, showing the power spectrum of p-modes sampled $20\text{ km}$ above the model surface. Blue dashed lines represent theoretical predictions of eigenmodes made by the standard solar model S \citep{1996Sci...272.1286C}.}
\label{fig:l-nu}
\end{figure}

\noindent The power spectrum of these modes can be visualized using an $l-\nu$ diagram, showing continuous radial modes throughout the solar interior as a function of their frequency ($\omega = 2\pi\nu$) and spherical harmonic degree ($l$), Fig. \ref{fig:l-nu}. The eigenmodes are excited in the frequency range determined by our source function ($S$, see \S\ref{sec:mathbg}). By setting negative values of the Brunt-V\"ais\"al\"a frequency to zero we remove the convective instabilities that act as sources of acoustic perturbations normally seen below $2\text{ mHz}$. We see an agreement in the structure of the eigenmodes generated by our model to theoretical calculations from the model S \citep{1996Sci...272.1286C}, denoted by by the dashed blue lines in Fig. \ref{fig:l-nu}.

\subsection{Rotation}\label{sec:rot}

\begin{figure}[h]
\center
\includegraphics[width=6cm]{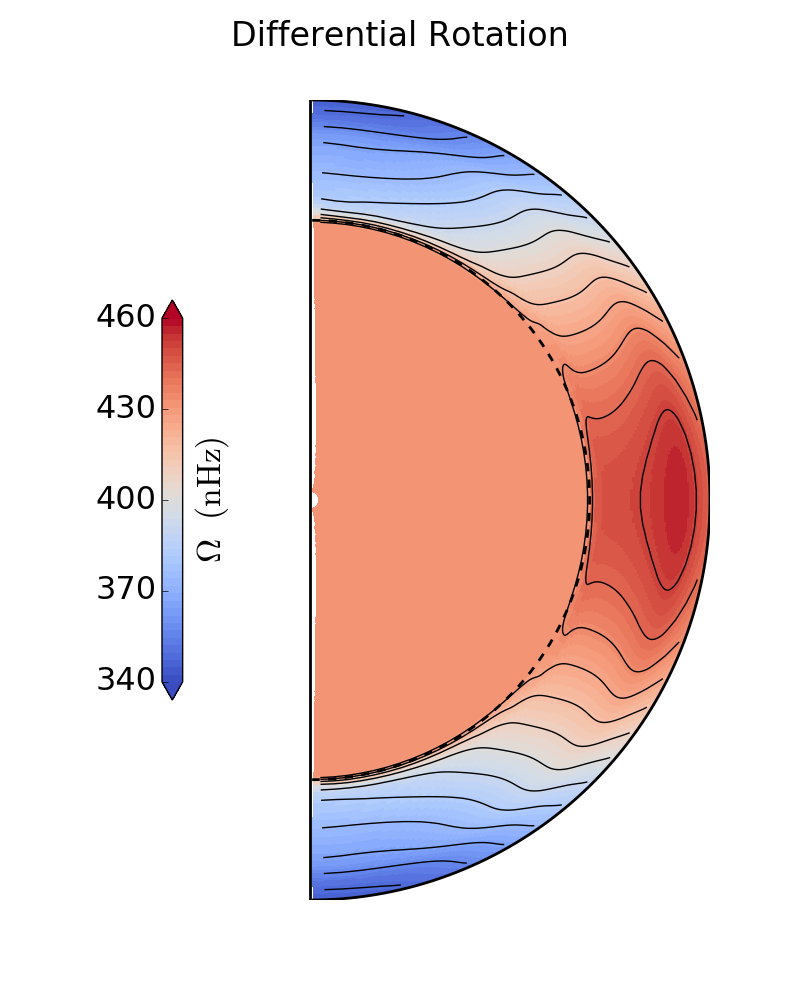}
\caption{The angular frequency ($\mathbf{\Omega}$) profile derived from model M1 described in \citet{2019ApJ...887..215P}, showing differential rotation in the convection zone ($0.70R_{\odot} - 1.001R_{\odot}$) with a solid rotating core ($<0.70R_{\odot}$).}
\label{fig:OM}
\end{figure}

\indent In order to test the computation of our material derivative (Eq. \ref{eq:gov2}), we implement a simple model of differential rotation in a non-rotating reference frame by defining our background velocity term as the angular frequency ($\mathbf{\Omega}$) derived from the mean-field model M1 described in \citet{2019ApJ...887..215P} and shown in Fig. \ref{fig:OM}.

\begin{figure}[h]
\center
\includegraphics[width=12cm]{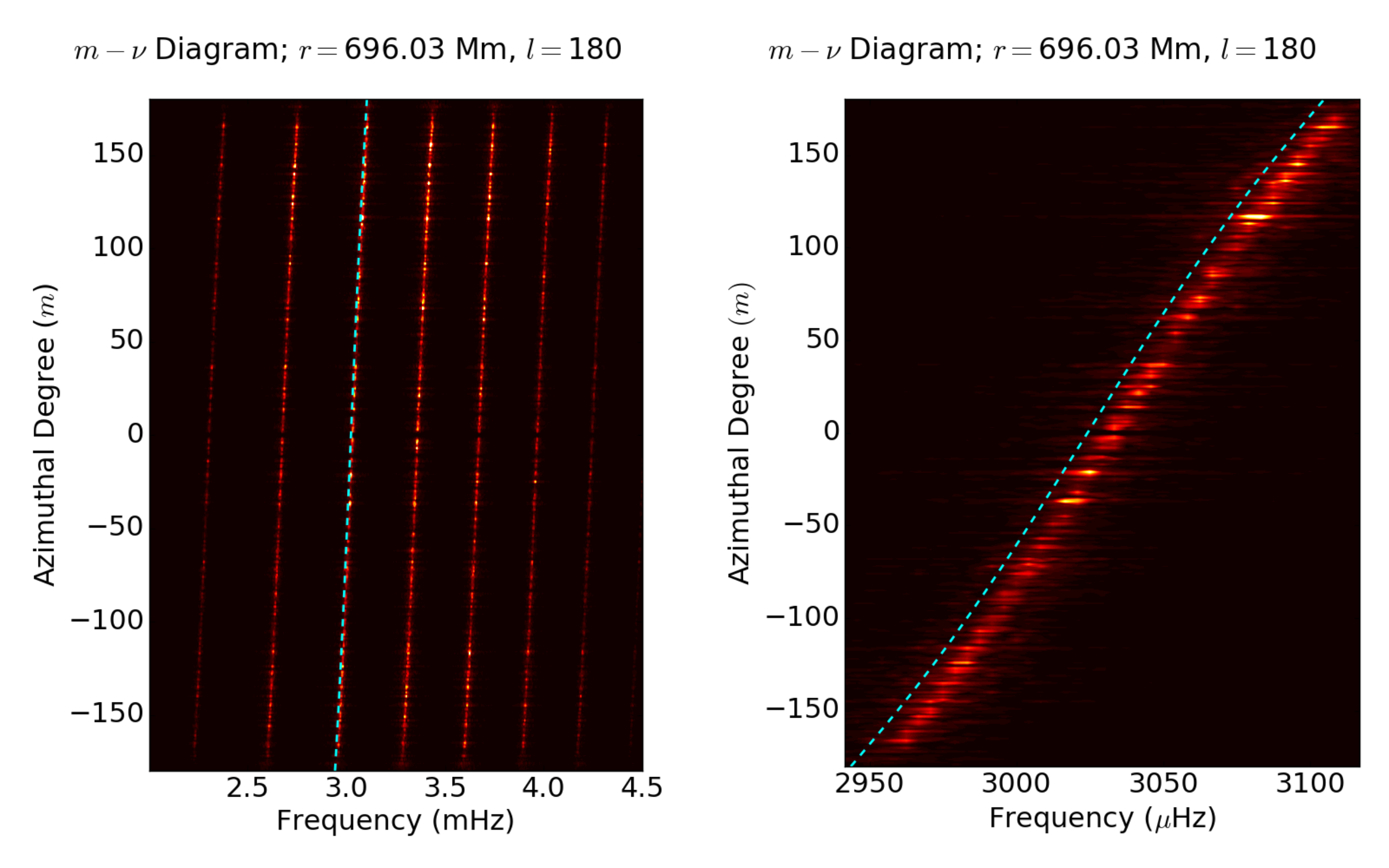}
\caption{An $m-\nu$ diagram, showing the power spectrum of acoustic oscillation sampled 30 km above the model surface, for spherical harmonic degree $l=180$. Blue dashed lines represent frequency splittings calculated using heliosesimic sensitivty kernels \citep{1998ApJ...505..390S} for the M1 model \citep{2019ApJ...887..215P}.}
\label{fig:m-nu}
\end{figure}

\noindent This simple azimuthal velocity flow field model creates easily detectable rotational splitting in the structure of our eigenmodes. The rotational profile will shift up the frequency of prograde modes and shift down the frequency of retrograde modes as a function of their azimuthal order ($m$). In order to visualize this shift we can use an $m-\nu$ diagram of the power spectrum for spherical harmonic degree $l=180$ (Fig. \ref{fig:m-nu}), where our simulated modes reproduce the characteristic tilt due to the average rotation along with the curvature created by the differential rotation in the convection zone. The blue dashed lines show frequency splittings calculated using heliosesimic sensitivity kernels \citep{1998ApJ...505..390S} for the M1 model \citep{2019ApJ...887..215P}.

\newpage

\section{Meridional Circulation}\label{sec:mer}

\begin{figure}[h]
\center
\includegraphics[width=6cm]{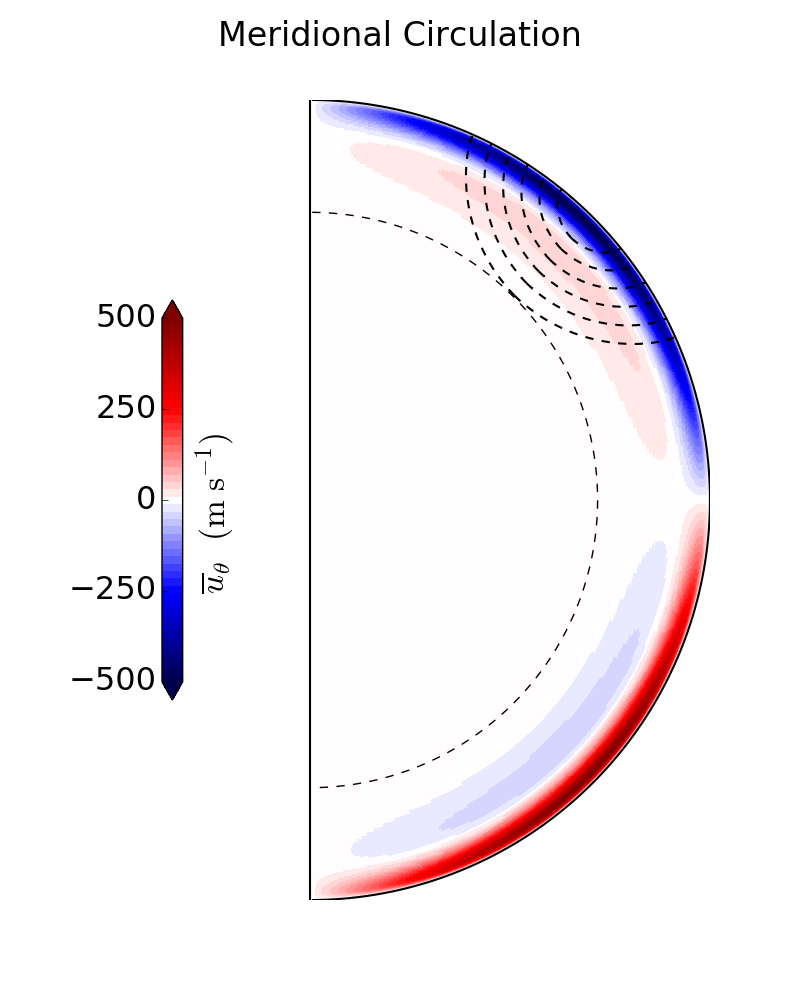}
\caption{The latitudinal velocity ($\tilde{u}_{\theta}$) of a single-cell model of meridional circulation \citep{2013ApJ...762..132H}. The dashed lines represent ray paths of acoustic oscillations (p-modes) between diameters of $\Delta = 12^{\circ}$, $18^{\circ}$, $24^{\circ}$, $30^{\circ}$, $36^{\circ}$ and $42^{\circ}$ with radial turning points at depths: $\sim 0.93$, $0.89$, $0.85$, $0.81$, $0.77$, $0.72\text{ R}_{\odot}$ respectively.}
\label{fig:MC}
\end{figure}

\indent The computation of our material derivative (Eq. \ref{eq:gov2}) can be tested in a regime of global meridional flows by using a simple single-cell model of meridional circulation as our background velocity term ($\tilde{\mathbf{u}}$, Fig. \ref{fig:MC}). We chose the model described and tested by \citet{2013ApJ...762..132H}, recreating their measurements as a simple validation of our computational techniques. \\ 
\indent In order to infer flows in the model interior using surface measurements, we apply a method of deep focusing \citep{2009ApJ...702.1150Z} -- a local helioseismology technique (for an in-depth review, see \citet{2005LRSP....2....6G}) which uses travel times of acoustic waves to probe structures in solar and stellar interiors. This method consists of choosing two points on the model surface, separated by some angular distance ($\Delta$); the signals at these points are cross-correlated, measuring acoustic travel times of internal oscillations (p-modes). As these waves travel through the resonant cavity of the convective interior they are advected by internal mass flows. Sampling waves traveling in opposite directions results in travel time differences ($\delta\tau$) which provide a basis for inferring background velocities. The ray path approximation under the assumption of Fermat's principle offers a simple relation between these values \citep{1999_giles}.

\begin{equation}\label{eq:raypath}
  \delta\tau = -2\int_{\Gamma_{0}} \dfrac{\tilde{\mathbf{u}}\cdot\mathbf{n}}{c^{2}} ds \ .
\end{equation}

\begin{figure}[!htb]
\center
\includegraphics[width=8cm]{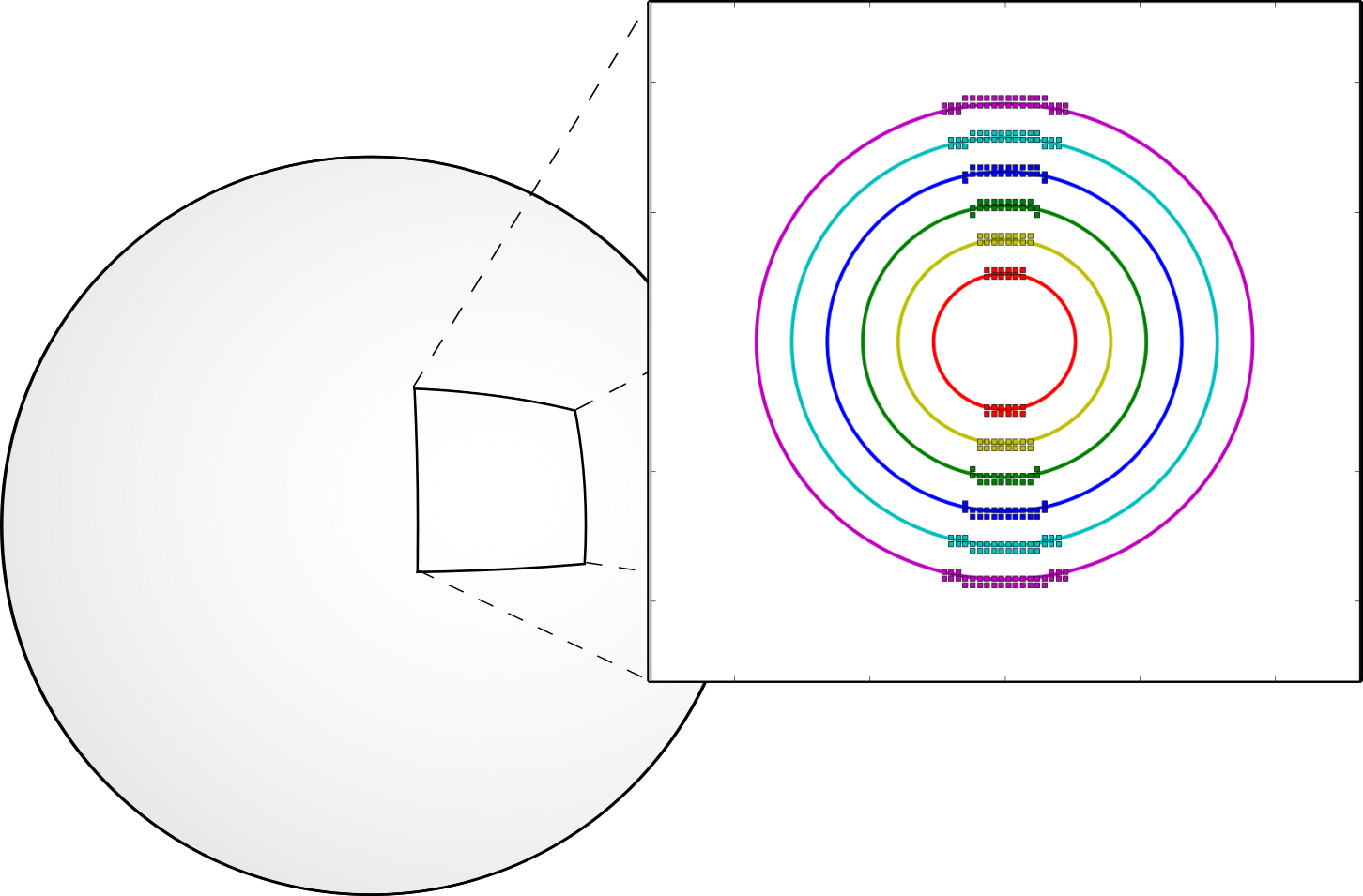}
\caption{An illustration of pixel selection for our deep-focusing method. A $60^{\circ}\text{ x }60^{\circ}$ patch is remapped into azimuthal equidistant coordinates to a resolution of approximately $0.6^{\circ}$ per pixel. Six concentric circles are selected at diameters $\Delta = 12^{\circ}$, $18^{\circ}$, $24^{\circ}$, $30^{\circ}$, $36^{\circ}$ and $42^{\circ}$. Pixels are chosen in $30^{\circ}$ wide northern and southern sectors two pixels in width.}
\label{fig:deepfoc}
\end{figure}

\noindent Travel time differences can be estimated along the unperturbed ray path ($\Gamma_{0}$). \\
\indent We employ this technique to infer meridional velocities by measuring travel time differences between southward and northward traveling waves ($\delta\tau_{NS}$). Using each pixel in our data set as a center point, we remap the surrounding $60^{\circ}\text{ x }60^{\circ}$ patch into azimuthal equidistant coordinates (Postel's proejection) using cubic hermite splines. The remapped resolution is approximately $0.6^{\circ}$ per pixel, the same spatial resolution as our Gauss-Legendre grid (\S\ref{sec:nume}). An illustration of our method can be seen in Fig. \ref{fig:deepfoc}.

\noindent A series of six concentric great circles are drawn at diameters of $\Delta = 12^{\circ}$, $18^{\circ}$, $24^{\circ}$, $30^{\circ}$, $36^{\circ}$ and $42^{\circ}$. Sets of pixels (two pixels in width) are selected along the rings in $30^{\circ}$ wide northern and southern sectors. The pixels in each sector are averaged together and the signals in opposing sectors are cross-correlated. This process is repeated for every grid point in our model ($N_{\phi}$, $N_{\theta}$) and the cross-correlated signal is averaged over every point in the longitude ($N_{\phi}$) and $\pm 3^{\circ}$ ($\pm 5\text{ px. in }N_{\theta}$) in the latitude. To further smooth our data, we average the diameter of each great circle over $\pm 2.4^{\circ}$ -- travel times of the varying diameter distances are interpolated to an estimated time offset based on ray path theory \citep{1999_giles}. The radial turning points of acoustic waves corresponding to each angular distance are $\sim 0.93$, $0.89$, $0.85$, $0.81$, $0.77$, $0.72\text{ R}_{\odot}$ respectively, allowing us to probe the entirety of the convective interior. In Fig. \ref{fig:MC} we show the meridional flow profile used in the model with dashed lines in the upper hemisphere corresponding to the ray paths for each great circle.

\subsection{Results}\label{sec:MCresults}

\begin{figure}[h]
\center
\includegraphics[width=16cm]{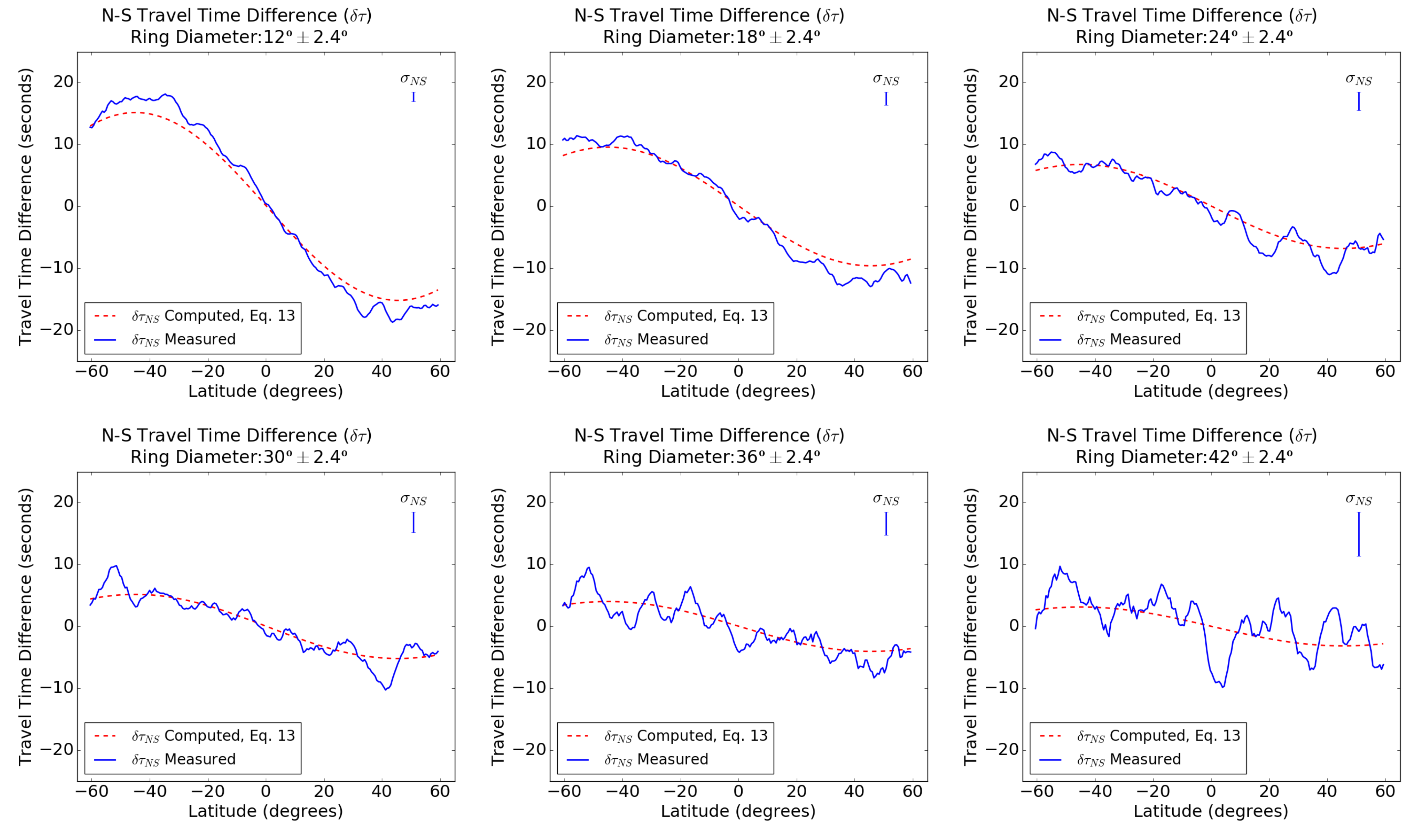}
\caption{The N-S travel-time differences ($\delta t_{NS}$) as a function of latitude for six depths: $\sim 0.93$, $0.89$, $0.85$, $0.81$, $0.77$, $0.72\text{ R}_{\odot}$ corresponding to travel distances of $\Delta = 12^{\circ}$, $18^{\circ}$, $24^{\circ}$, $30^{\circ}$, $36^{\circ}$ and $42^{\circ}$ respectively. The signal is averaged over $\pm 3^\circ$ in latitude and $\pm 2.4^\circ$ in travel distance.}
\label{fig:dt_NS}
\end{figure}

\indent The travel time differences ($\delta\tau_{NS}$) for each ring diameter are plotted as a function of latitude in Fig. \ref{fig:dt_NS}. These values are compared to theoretical travel time differences (dashed lines in Fig. \ref{fig:dt_NS}) computed using the ray path approximation (Eq. \ref{eq:raypath}) employing the standard solar model S \citep{1996Sci...272.1286C}.

\indent The travel time differences for all ring diameters show solid agreement with theoretical predictions as well as the analysis of \citet{2013ApJ...762..132H}. These results show a key validation of the numerical procedure used to compute the model as well as the deep focusing techniques used to analyze the data. The error ($\sigma_{NS}$) is calculated using a separate model with no background flows; this reference model uses an identical source function ($S$, Eq. \ref{eq:gov2}) and analysis sequence, producing the same error profile we see for each ring diameter (Fig. \ref{fig:dt_NS}). We characterize this error as the standard deviation of travel time differences ($\delta\tau$) from zero in our reference model, taking the RMS over latitudinal grid points.

\begin{equation}\label{eq:error}
  \sigma_{NS} = \sqrt{\dfrac{1}{N}\sum_{i=1}^{N}  \delta\tau_{i}^{2}}\ .
\end{equation}

\noindent The characteristic noise profiles seen in models without flows can be subtracted from measured travel-time differences (Fig. \ref{fig:dt_NS}) in order to remove some of the most significant impacts of realization noise on our measurements \citep{2007ApJ...664.1234H}. This method of attenuating noise can provide us an opportunity to compare measurements made by computational helioseimology to estimates of the ray-path approximation. The resulting NS travel-time profile can be found in Appendix \ref{sec:rnoise}, Fig. \ref{fig:dt_NSU_rm}.

\begin{figure}[h]
\center
\includegraphics[width=16cm]{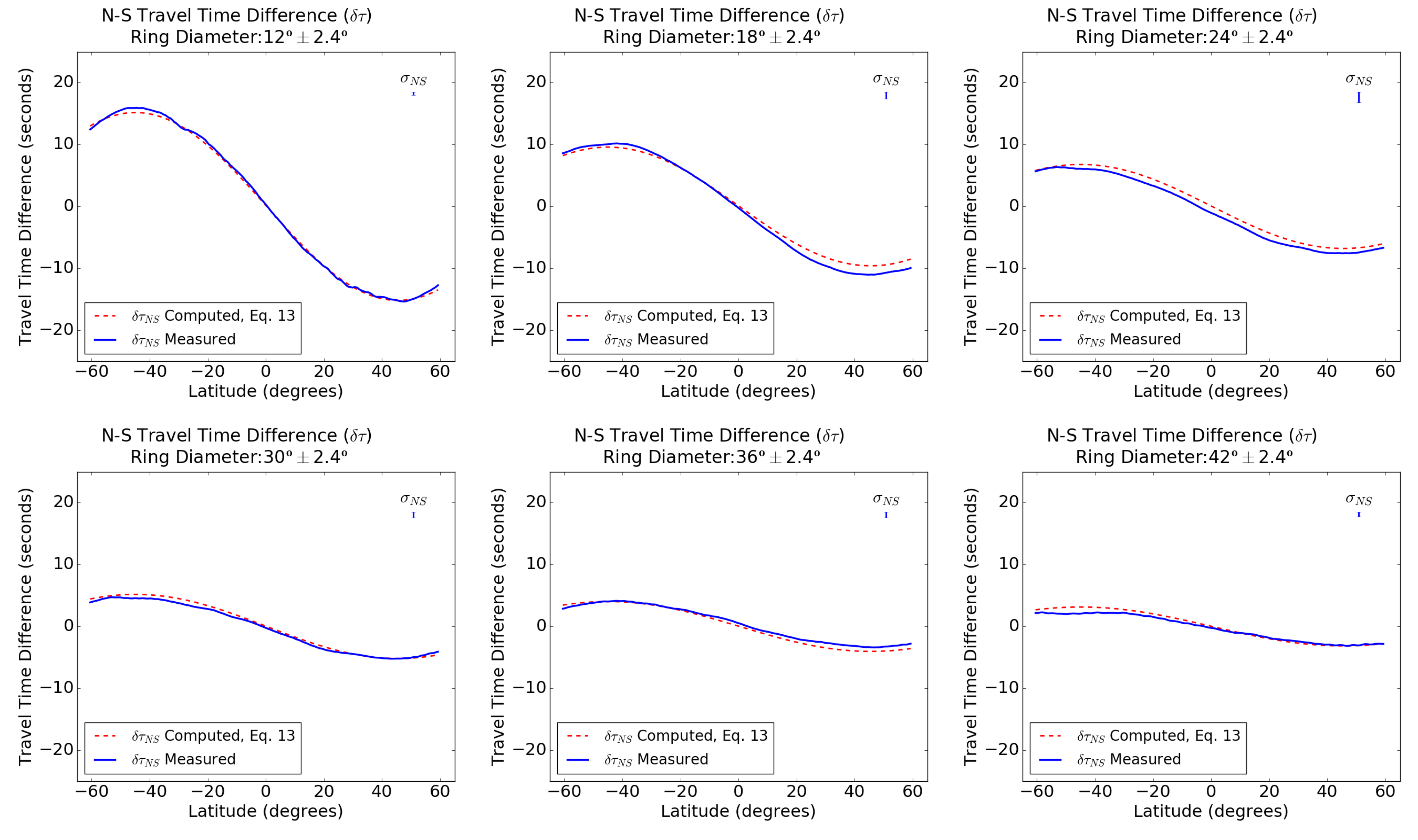}
\caption{The N-S travel-time differences ($\delta t_{NS}$) under the application of a Gaussian phase speed filter ($\sigma = 0.05v_{p}$) as a function of latitude for six depths: $\sim 0.93$, $0.89$, $0.85$, $0.81$, $0.77$, $0.72\text{ R}_{\odot}$ corresponding to travel distances of $\Delta = 12^{\circ}$, $18^{\circ}$, $24^{\circ}$, $30^{\circ}$, $36^{\circ}$ and $42^{\circ}$ respectively. The signal is averaged over $\pm 3^\circ$ in latitude and $\pm 2.4^\circ$ in travel distance.}
\label{fig:dt_NS_F}
\end{figure}

\indent In order to further increase the SNR in our measurements we apply a phase speed filter, defined by a Gaussian function with a width of $\sigma = 0.05 v_{p}$, where $v_{p} = \omega/L$ is the phase speed \citep{2007ApJ...659.1736N}. After the application of this filter, the travel time differences display high levels precision but do show a latitude-independent systematic offset for different ring diameters. This offset, which seems to have a maximum of approximately $\pm 1\text{ s}$ (Fig. \ref{fig:dt_NS_F}), is a characteristic of the realization noise in our model and can be removed by subtracting the phase-filtered noise profile for the same model with no background flows. The resulting profiles, which can be found in Appendix \ref{sec:rnoise}, Fig. \ref{fig:dt_NS5_rm}, show remarkable agreement with predictions made using the ray-path approximation (Eq. \ref{eq:raypath}), implying that non-linear effects do not seem to have dramatic impacts on our measurements of travel times.

\begin{figure}[h]
\center
\includegraphics[width=14cm]{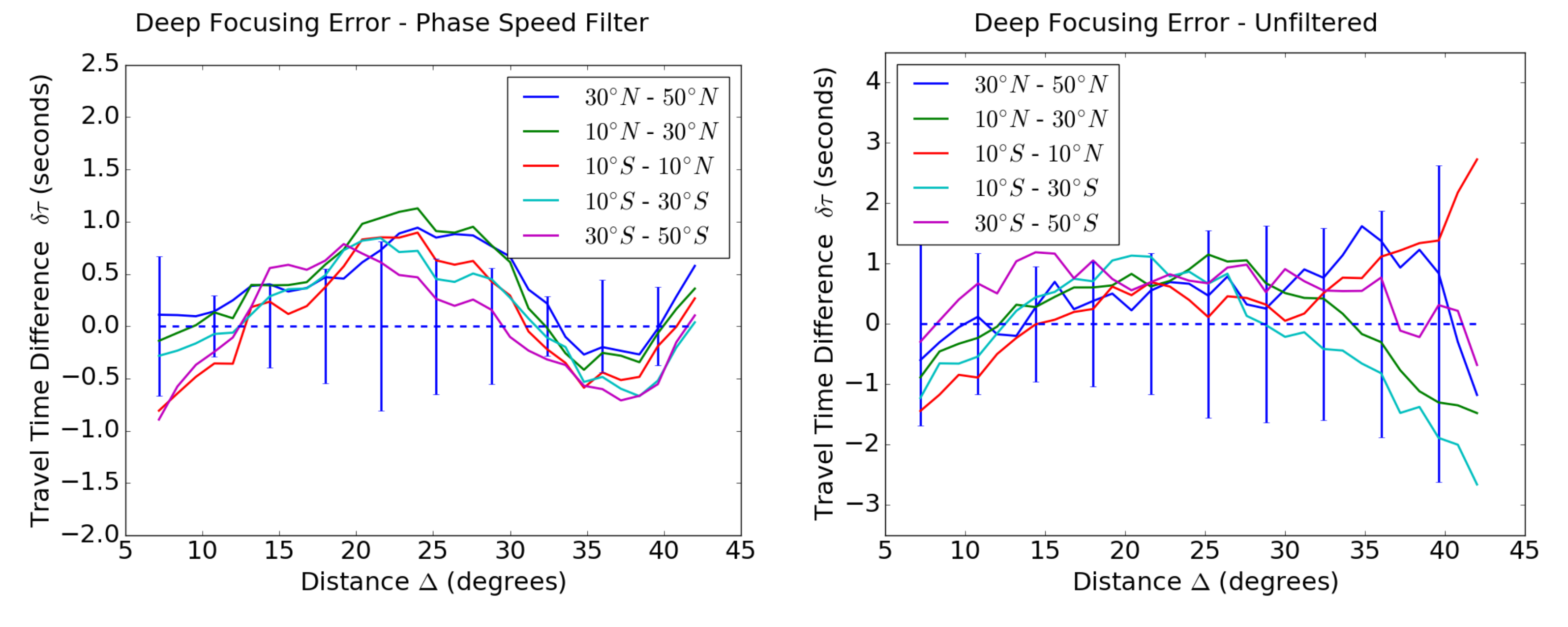}
\caption{The error in travel time differences ($\delta\tau$) as a function of travel distance ($\Delta$) for 5 latitudinal averages spanning $30^{\circ}\text{N} - 50^{\circ}\text{N}$, $10^{\circ}\text{N} - 30^{\circ}\text{N}$, $10^{\circ}\text{S} - 10^{\circ}\text{N}$, $10^{\circ}\text{S} - 30^{\circ}\text{S}$, $30^{\circ}\text{S} - 50^{\circ}\text{S}$. Error bars show the standard deviation of the measured offset ($\sigma_{NS}$, Eq. \ref{eq:error}) across the entire latitude. Left) Error for data analyzed with a Gaussian phase speed filter ($\sigma = 0.05v_{p}$). Right) Error for unfiltered signal.}
\label{fig:TD_err}
\end{figure}

\indent The error function ($\sigma_{NS}$, Eq. \ref{eq:error}) provides a solid foundation for the characterization of realization noise at various acoustic travel depths. We show the error as a function of travel distance ($\Delta$) in Fig. \ref{fig:TD_err}, along with travel time differences ($\delta\tau$) from our reference model without background flows for 5 separate latitudinal averages ($30^{\circ}\text{N} - 50^{\circ}\text{N}$, $10^{\circ}\text{N} - 30^{\circ}\text{N}$, $10^{\circ}\text{S} - 10^{\circ}\text{N}$, $10^{\circ}\text{S} - 30^{\circ}\text{S}$, $30^{\circ}\text{S} - 50^{\circ}\text{S}$) and compare them with the error in the unfiltered signal (Fig. \ref{fig:dt_NS}). The cause of the apparent systematic error is unclear, however, the latitude-dependent offset profiles are strongly linked to the structure of our source ($S$, Eq. \ref{eq:gov2}), with different random number seeds generating different error profiles. This effect deserves its own systematic investigation for varying parameters of source locations and structures.

\indent In Fig. \ref{fig:TD_err}, we see a similar systematic error structure in both the filtered and unfiltered signals. As we move towards greater depths, however, noise in the unfiltered signal grows significantly, concealing any potential offset. These results may have interesting implications for measuring meridional flow structures at the base of the tachocline. The application of our phase speed filter seems to have preserved signal quality relatively evenly throughout the convection zone, offering encouraging results for probing flows deep in the solar interior.

\section{Discussion}\label{sec:conclusions}

\indent We present a global linearized acoustic algorithm with new computational methods that will facilitate the testing and validation of local and global helioseismology techniques in diverse regimes of 3-dimensional flows. While helioseismology has been an indispensable tool in exploring interior dynamics on the Sun, it can have trouble resolving exact profiles of flow, especially at greater depths. Forward modeling offers the opportunity to test the impact of subtle differences generated by a variety of theoretical models of mean mass flows, forming a basis to better interpret observational oscillation data.\\ 
\indent The model is validated for two distinct profiles: differential rotation and meridional circulation. These two regimes are critical for understanding and simulating the distribution of angular momentum and magnetic flux that governs the solar magnetic cycle. To simulate these structures, we present an application of pseudo-spectral techniques which will be necessary components for future models to efficiently compute spherical harmonic resolutions of $l_{max}>300$, previously considered to be too computationally expensive. These resolutions will be necessary to model local helioseismology techniques on sound speed perturbations due to small-scale structures on the solar surface, creating a link to linear effects of global perturbations.\\
\indent In future work, we plan to use this model to test differences in acoustic travel-time signatures between models of one and two-cell meridional circulation. The current state surrounding the nature of meridional velocity profiles remains uncertain, with similar techniques producing widely varying results between observational data from HMI \citep{2013ApJ...774L..29Z} and MDI/GONG \citep{2020Sci...368.1469G}. While both studies employed validation through the use of forward-modeling, the theoretical profiles of meridional circulation structures for one and two-cell models varied extremely. A new efficient and flexible code presents a timely opportunity to run a differential study on much more subtle differences between one and two-cell circulation models -- for both theoretical profiles as well as ones derived from solar simulations. Our investigation (Stejko, Kosovichev, in preparation) will explore subtle differences in flow velocities near the base of the tachocline, focusing on the limits what can be resolved within the two-decade long window of current solar observations, with and without preparing data using phase-speed filtering. These models will be used for an analysis of realization noise in simulations of stochastic excitations of surface oscillations -- for large numbers source profiles and measurement times. This level of systematic investigation will provide a baseline for what can be interpreted from observational data. In further work, we also plan to create a new generation of models that include linear effects of magnetic field structures on acoustic oscillations. The numerical method demonstrated in this paper will be the basis for computing this effect.

\acknowledgments

\indent AMS would like to thank the heliophysics modeling and simulation team at NASA Ames Research Center for very productive meetings. This work is supported by the NASA grants: 80NSSC19K0630, 80NSSC19K1436, NNX14AB7CG and NNX17AE76A.

\appendix
\section{Energy equation}\label{sec:eeq}

\indent In our formulation, we employ the adiabatic approximation, where we assume that heat transfer is negligible in the time-scale of acoustic oscillations. The relationship between pressure and density in this regime (see \citet{christensen14}) can be expressed in Eulerian form as

\begin{equation}
\label{eq:ecade}
  \dfrac{\partial p}{\partial t} + \mathbf{u}\cdot\nabla p = \dfrac{\Gamma_{1}p}{\rho}\left(\dfrac{\partial \rho}{\partial t} + \mathbf{u}\cdot\nabla\rho\right) \ .
\end{equation}

\noindent Assuming that the adiabatic ratio ($\Gamma_{1}$) remains constant, we can linearize Eq. (\ref{eq:ecade}) as follows, where primes denote a perturbation from a base (tilde) value.

\begin{multline}\label{eq:ecp}
  \tilde{\rho}\dfrac{\partial p'}{\partial t} + \rho'\dfrac{\partial\tilde{p}}{\partial t} + \left(\rho'\tilde{\mathbf{u}}\cdot\nabla\tilde{p} + \tilde{\rho}\left(\tilde{\mathbf{u}}\cdot\nabla p' + \mathbf{u}'\cdot\nabla\tilde{p}\right)\right) = \Gamma_{1}\left(\tilde{p}\dfrac{\partial \rho'}{\partial t} + p'\dfrac{\partial\tilde{\rho}}{\partial t} + \left(p'\tilde{\mathbf{u}}\cdot\nabla\tilde{\rho} + \tilde{p}\left(\tilde{\mathbf{u}}\cdot\nabla\rho' + \mathbf{u}'\cdot\nabla\tilde{\rho}\right)\right)\right) \ .
\end{multline}

\noindent In order to avoid convective instabilities in our model we can rewrite Eq. (\ref{eq:ecp}) in terms of the Brunt-V\"ais\"al\"a frequency ($N^{2}$, Eq. \ref{eq:BVf}). By plugging in Continuity (Eq. \ref{eq:gov1}) and rearranging terms, we are left with:

\begin{equation}
  \dfrac{1}{\tilde{p}\Gamma_{1}}\dfrac{\partial p'}{\partial t} + \dfrac{1}{\tilde{\rho}}\nabla\cdot\left(\tilde{\rho}\mathbf{u}'+\rho'\tilde{\mathbf{u}} \right) = - \mathbf{u}'\cdot\left(\dfrac{1}{\Gamma_{1}\tilde{p}}\nabla\tilde{p} - \dfrac{1}{\tilde{\rho}}\nabla\tilde{\rho} \right) - \tilde{\mathbf{u}}\cdot\left( - \dfrac{1}{\tilde{\rho}}\left(\nabla\rho' + \dfrac{p'}{\tilde{p}}\nabla\tilde{\rho}\right)\right) \ .
\end{equation}

\noindent Substituting ($N^{2}$, Eq. \ref{eq:BVf}) and rearranging terms further leaves us with the final form of our governing relation for pressure (Eq. \ref{eq:gov3}).

\begin{equation}
  \dfrac{\partial p'}{\partial t} = - \dfrac{\Gamma_{1}\tilde{p}}{\tilde{\rho}}\left(\nabla\cdot\tilde{\rho}\mathbf{u}'+ \rho'\nabla\cdot\tilde{\mathbf{u}} - 
  \dfrac{p'}{\tilde{p}}\tilde{\mathbf{u}}\cdot\nabla\tilde{\rho} + \tilde{\rho} \mathbf{u}'\cdot\dfrac{N^{2}}{g}\mathbf{\hat{r}}\right) \ .
\end{equation}

\newpage

\section{Removing Noise}\label{sec:rnoise}

\begin{figure}[H]
\center
\includegraphics[width=15.5cm]{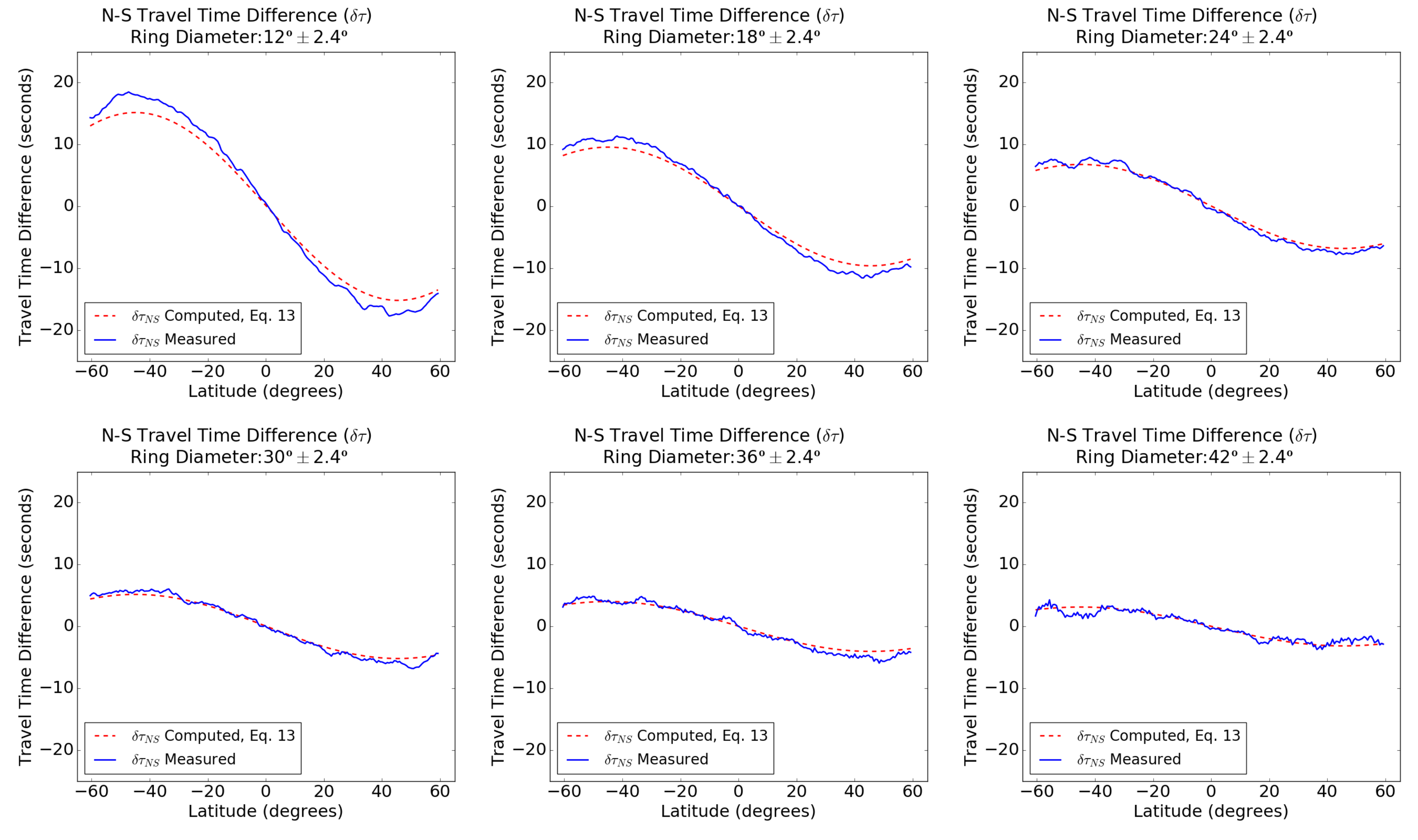}
\caption{The N-S travel-time differences ($\delta t_{NS}$) with noise subtracted from corresponding model with no flows. The travel-times are plotted as a function of latitude for six depths: $\sim 0.93$, $0.89$, $0.85$, $0.81$, $0.77$, $0.72\text{ R}_{\odot}$ corresponding to travel distances of $\Delta = 12^{\circ}$, $18^{\circ}$, $24^{\circ}$, $30^{\circ}$, $36^{\circ}$ and $42^{\circ}$ respectively. The signal is averaged over $\pm 3^\circ$ in latitude and $\pm 2.4^\circ$ in travel distance.}
\label{fig:dt_NSU_rm}
\end{figure}

\vskip-0.5cm

\begin{figure}[H]
\center
\includegraphics[width=15.5cm]{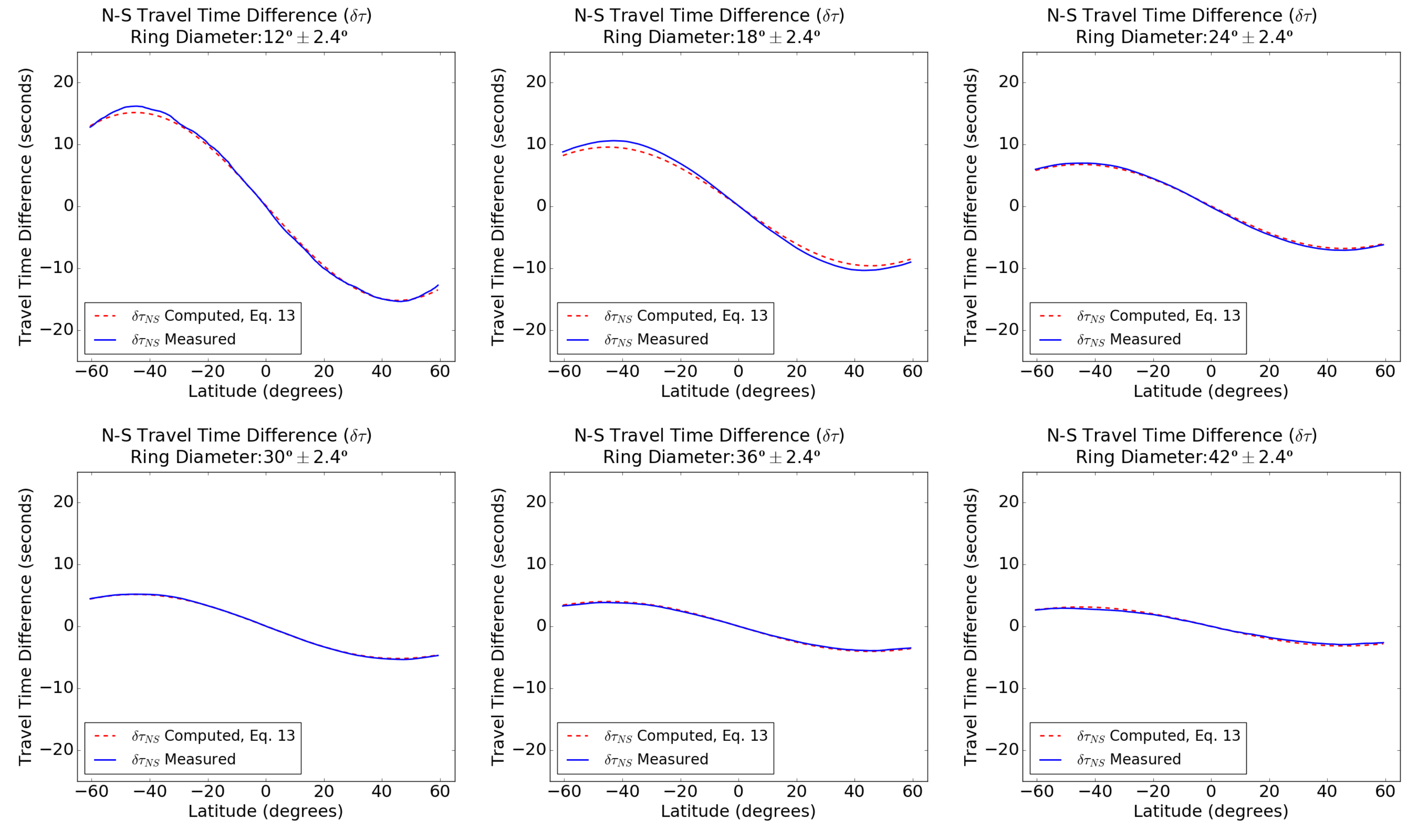}
\caption{The N-S travel-time differences ($\delta t_{NS}$) under the application of a Gaussian phase speed filter ($\sigma = 0.05v_{p}$) with noise subtracted from corresponding model with no flows. The travel-times are plotted as as a function of latitude for six depths: $\sim 0.93$, $0.89$, $0.85$, $0.81$, $0.77$, $0.72\text{ R}_{\odot}$ corresponding to travel distances of $\Delta = 12^{\circ}$, $18^{\circ}$, $24^{\circ}$, $30^{\circ}$, $36^{\circ}$ and $42^{\circ}$ respectively. The signal is averaged over $\pm 3^\circ$ in latitude and $\pm 2.4^\circ$ in travel distance.}
\label{fig:dt_NS5_rm}
\end{figure}



\end{document}